\documentclass[%
 aapm,
 mph,%
 amsmath,amssymb,
 reprint,%
]{revtex4-2}

\usepackage{graphicx}
\usepackage{dcolumn}
\usepackage{bm}
\usepackage{color}

\usepackage[mathlines]{lineno}
\modulolinenumbers[5]

\begin{document}

\preprint{AAPM/123-QED}

\title[Ghost channels and cycles] {Ghost channels and ghost cycles guiding long transients in dynamical systems}

\author{D. Koch$^{1,\dagger}$}
\author{A. Nandan$^{1,\dagger}$}
\author{G. Ramesan$^{1}$}
\author{I. Tyukin$^{2}$}
\author{A. Gorban$^{3}$}
\author{A. Koseska$^{1,\ddagger}$}
\affiliation{$^1$ Cellular computations and learning, Max Planck Institute for Neurobiology of Behavior – caesar, Bonn, Germany}
\affiliation{$^2$ Department of Mathematics, King's College, London, UK}
\affiliation{$^3$Department of Mathematics, University of Leicester, Leicester, UK}

\thanks{$^{\dagger}$Equally contributing authors;\\
$^{\ddagger}$Corresponding author:aneta.koseska@mpinb.mpg.de}

\date{\today}
\begin{abstract}
Dynamical descriptions and modeling of natural systems have generally focused on fixed points, with saddles and saddle-based phase-space objects such as heteroclinic channels/cycles being central concepts behind the emergence of quasi-stable long transients. Reliable and robust transient dynamics observed for real, inherently noisy systems is, however, not met by saddle-based dynamics, as demonstrated here. Generalizing the notion of ghost states, we provide a complementary framework that does not rely on the precise knowledge or existence of (un)stable fixed points, but rather on slow directed flows organized by ghost sets in {\textit{ghost channels}} and {\textit{ghost cycles}}. Moreover, we show that appearance of these novel objects is an emergent property of a broad class of models, typically used for description of natural systems.

\end{abstract}

\maketitle
\begin{figure}
\includegraphics[scale=1]{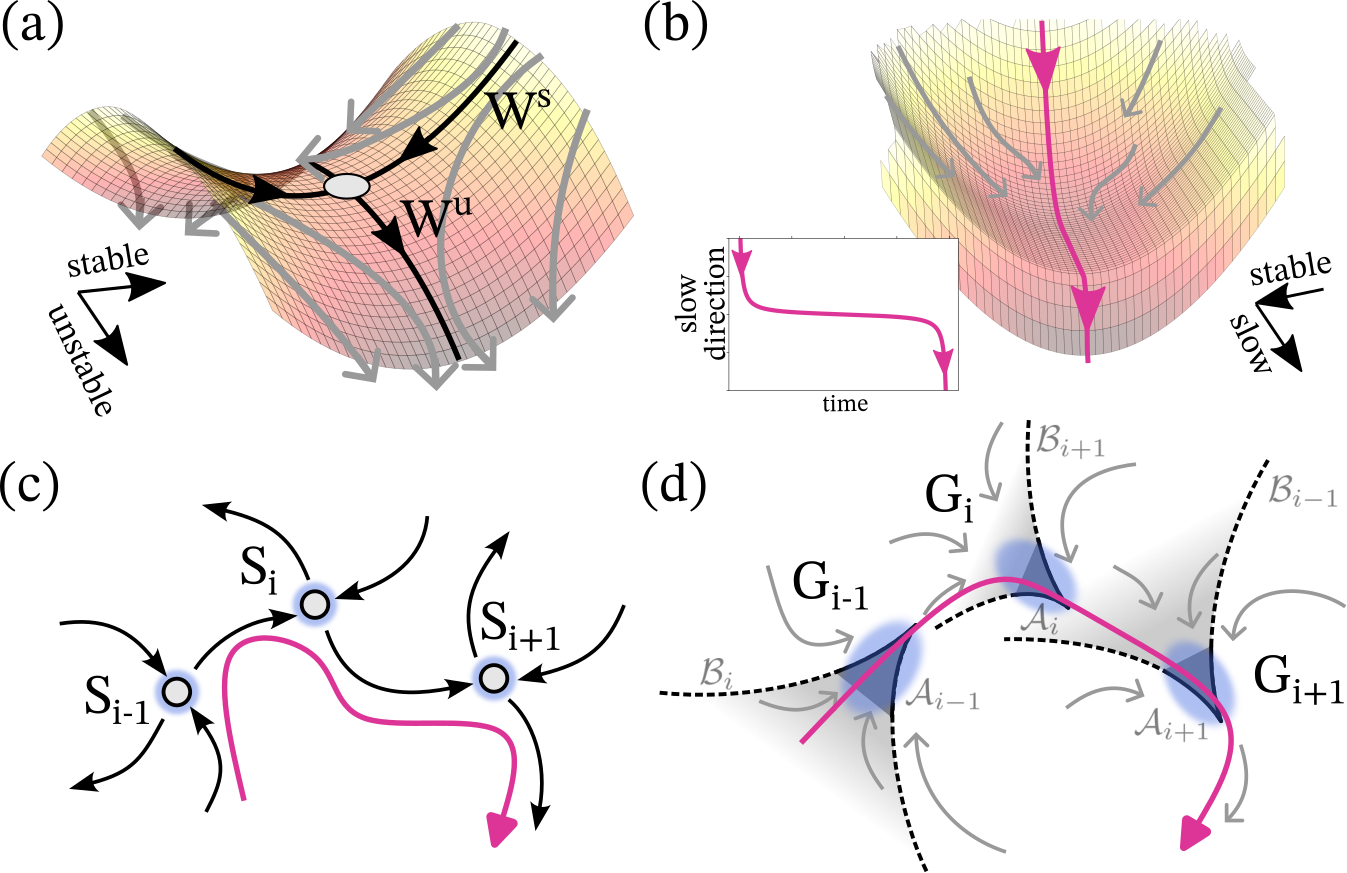}
\caption{Schematics of phase-space objects. (a) Quasi-potential landscape of a saddle fixed point. Gray dot: unstable fixed point localization. (b) Quasi-potential landscape of a ghost state. Note the absence of a fixed point. Inset: time-course of a trajectory with slow transition through the set. (c) Schematic diagrams of scaffolds of connected (c) saddles ($S_i$), i.e. heteroclinic channel, and (d) ghosts ($G_i$), i.e. ghost channel. $A_i$ denotes the ghost-attracting set of $G_i$, and $B_i$ its basin. (a-d) Black, gray and magenta arrows represent (un)stable manifolds, flow direction and example trajectories, respectively.}
\label{fig:fig1}
\end{figure}

Living and man-made, but also ecological or climate systems are classically described to reproduce asymptotic behavior, implying that the observed dynamics is retained indefinitely in absence of a perturbation. Mathematically, such dynamics corresponds to invariant sets that represent objects in phase-space, the simplest being stable fixed points separated by separatrixes of saddles \cite{Guckenheimer_1983,strogatz:1994}. However, a growing body of empirical evidence suggests that real-world systems are often characterized by long transients which are not invariant, are quasi-stable, and the system switches between them. The duration of the quasi-stable patterns is much longer than one would expect from the characteristic elementary processes, whereas the switching is triggered by external signals or system-autonomously, and occurs on a timescale much shorter than the one of the preceding dynamical pattern. Examples include dynamics of neuronal activity\cite{Mazor_2005,Benozzo_2021,Recanatesi_2022}, camouflaging in animals\cite{Woo_2023}, cell signaling \cite{Nandan_2022,Karin_2023,Tufcea_2015}, ecological \cite{Hastings_2018,Bieg_2022}, earth and climate systems \cite{Colin_de_Verdi_re_2007,Kasz_s_2019}, replicator networks \cite{Sardany_s_2007}, semiconductor lasers and Josephson junctions \cite{Strogatz_1989}. In the context of neuronal systems, the described dynamics is often referred to as \textit{metastable} \cite{Kelso2012, Deco2016, Rossi2023,Bovier_2015}. 
From the aspect of topological dynamics, long transients are caused by the bifurcations (explosions) of $\omega$-limit sets or through the occurrence of generalized homoclinic (loop) structures \cite{Gorban_1980,Gorban_1981,Gorban_2004}. In particular, the emergence of long transients has been conceptualized by trapping the system's dynamics in the vicinity of a saddle\cite{Hastings_2018} (Fig.~\ref{fig:fig1}(a)), or a saddle-node `ghost'\cite{Gorban_2004, Hastings_2018, Nandan_2022, Karin_2023} (Fig.~\ref{fig:fig1}(b)), whereas the switching is thought to occur via saddle-based heteroclinic structures\cite{May_1975,Rabinovich_2001, Kirst_2008,Ashwin_2013,Horchler_2015} (Fig.~\ref{fig:fig1}(c)).

In this Letter, we complement the topological theory with novel structures. Generalizing the concept of ghost states\cite{Gorban_2004,Strogatz_1989,strogatz:1994}, we provide a theoretical framework for generation of sequential quasi-stable dynamics that does not rely on (un)stable fixed points, but on slow directed phase-space flows guided by {\textit {ghost channels}} and ghost cycles (Fig.~\ref{fig:fig1}(d)). We identify the criteria for their emergence, and demonstrate that ghost-based scaffolds capture properties of long transients in noisy systems better than traditional models relying on invariant sets.

\textit{Properties of ghost sets.}\textemdash To formally define ghost channels and cycles, we need to provide a quantitative description of ghost sets, their boundaries and trapping times. Consider a conceptual 2D-system of first order differential equations $\dot{\bf{x}}=\bf{F(x)}$ given by

\begin{equation}
    \qquad \dot{x} = \alpha+x^2, \qquad \dot{y} = -y.
\end{equation}
For $\alpha<0$, a stable fixed point and a dissipative saddle coexist, whereas for $\alpha \rightarrow 0^{+}$, a ghost state or a `bottleneck' appears\cite{strogatz:1994} (Supplementary Fig.1(a), numerical analysis as in Ref. [27]).
 
\begin{figure}[!h]
\includegraphics[width=0.5\textwidth]{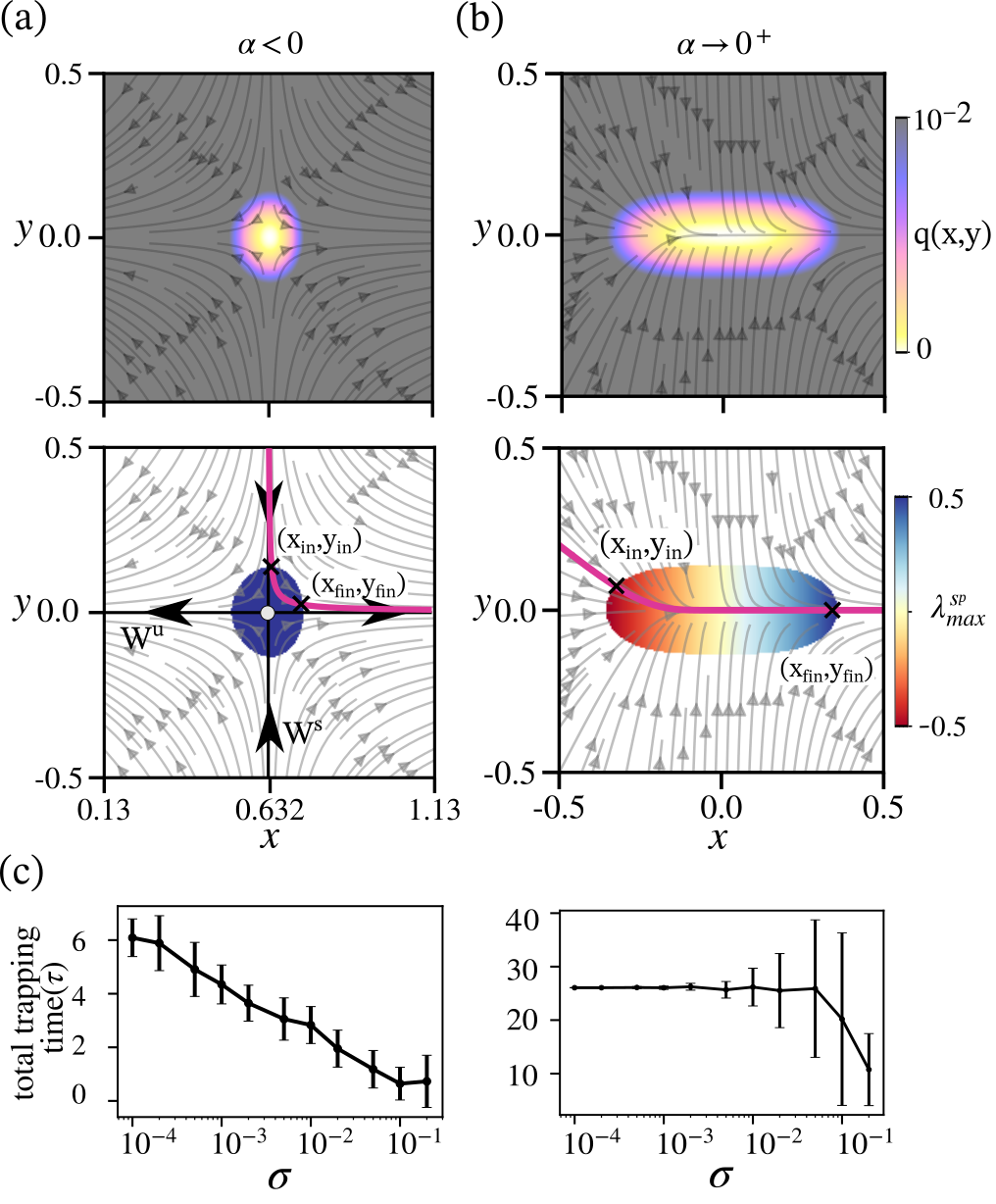}
\caption{Characteristics of quasi-stable transients emerging from saddles vs. ghost states. (a) Top: Kinetic energy estimate ($q(x,y)$) around a saddle fixed point ($\alpha=-0.4$ in Eqs.1); Bottom: corresponding maximum eigenvalue ($\lambda_{\text{max}}^{sp}$) in the $q(x^{sp},y^{sp})<q_{\text{thresh}}=0.01$ region; Stable/unstable (\textbf{$W^s$},\textbf{$W^u$}) manifold (black arrow lines) and exemplary phase-space trajectory (magenta line). Black crosses: entry/exit points; (b) Same as in (a), but for ghost state ($\alpha=0.01$ in Eqs.1). Gray arrow lines: phase-space flow. 
(c) Dependence of the total trapping time in the $q(x^{sp},y^{sp})<q_{\text{thresh}}$ of the saddle (left) and the ghost (right) for different additive noise intensities $\sigma$. Mean $\pm$ s.d. from 30 repetitions are shown. See Supplementary materials.
}
\label{fig:fig2}
\end{figure}

Fixed points in dynamical systems (and their surrounding regions of slow dynamics) can be identified using an auxiliary scalar function $q({\bf x})=\frac{1}{2}|\bf{F(x)}|^2$, which is also a Lyapunov function, related to the system's kinetic energy\cite{Sussillo2013} where $|.|$ is the modulus operator. Calculating the kinetic energy for system Eqs.1 shows that $q({\bf{x^{*}}})=0$ if and only if $x^{*}$ is a fixed point, i.e. the saddle (Fig.~\ref{fig:fig2}(a), top; the stable fixed point is omitted for brevity). In its proximity $q$ adopts values close to $0$ ($q_{\text{thresh}}=10^{-2}$), corresponding to the upper surface of the saddle (Fig.\ref{fig:fig1}(a)). Slow dynamics with $q<q_{thresh}$ still occurs for $\alpha \rightarrow 0^{+}$ , spanning across an even larger phase-space area (Fig.~\ref{fig:fig2}(b), top). This area corresponds to the shallow-slope region of the ghost state in Fig.\ref{fig:fig1}(b).
If the norm of the dynamics is close to zero ($q_{\text{thresh}}=10^{-2}$), local linear expansion is still valid \cite{Sussillo2013}. Numerically evaluating the two local eigenvalues, $\lambda_{\text{max}}^{sp}$ and $\lambda_{\text{min}}^{sp}$, for every slow point $(x^{sp},y^{sp})$ satisfying $q(x^{sp},y^{sp})<q_{\text{thresh}}$ using the Jacobian of Eqs.1 shows that $\lambda^{sp}_{min}$ is negative for $\alpha<0$ and $\alpha \rightarrow 0^{+}$ in the whole slow region (Supplementary section I, Supplementary Fig.1(b)). $\lambda^{sp}_{\text{max}}$, however, remains positive around the saddle as characteristic for unstable fixed points, while for $\alpha \rightarrow 0^{+}$ the $\lambda^{sp}_{\text{max}}$ changes from negative to positive in the slow-dynamics region  (Fig.~\ref{fig:fig2}(a-b), bottom). 
The corresponding eigenvectors thereby determine the flow direction: trajectories starting along the stable saddle manifold are deflected along its unstable manifold, whereas for $\alpha \rightarrow 0^{+}$ the phase-space flow attracts and guides the trajectories along the low kinetic energy area (Fig.~\ref{fig:fig2}(a-b), bottom, respectively). Thus, the system's trajectories, although transiently attracted to the ghost state, eventually escape. This is also contrary to a saddle-node (i.e. when saddle and stable fixed point collide at $\alpha=0$), as the trajectories coming from the left will be trapped at the origin. 

\begin{figure}
\includegraphics[scale=1]{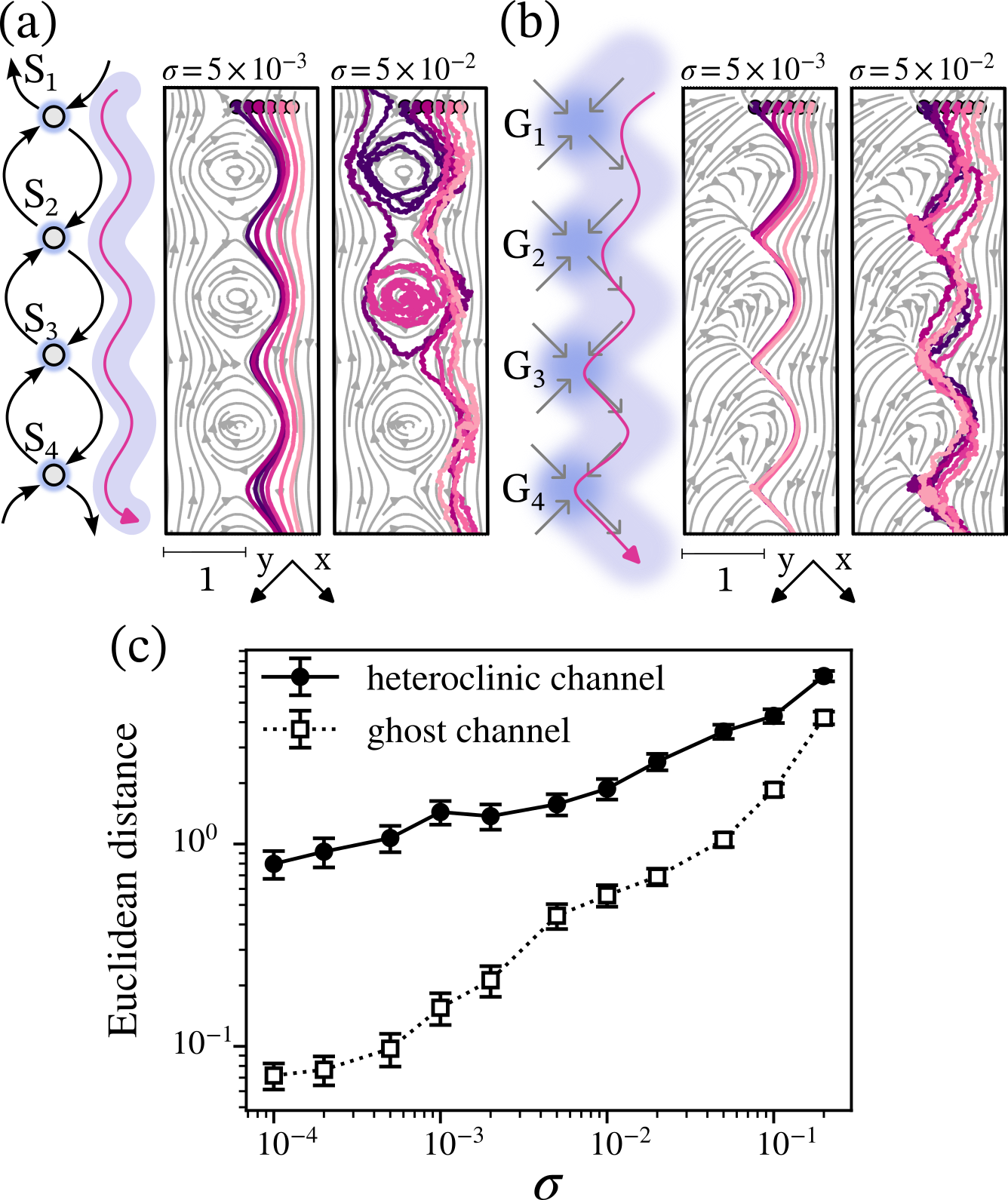}
\caption{\label{fig:fig3}Comparison of heteroclinic (HCh) and ghost (GCh) channels' dynamics. (a) Schematic of a HCh and exemplary trajectories for six initial conditions and two different noise intensities $\sigma$. (b) Same as in (a), but for a GCh. (c) Euclidean distance between pairwise trajectories in the HCh or GCh as a function of $\sigma$ (mean $\pm$ SEM from 180 trajectories: six initial conditions with 30 repetitions; Supplementary materials).}
\end{figure}

Since the trajectories in the ghost travel a along phase-space region where $\lambda^{sp}_{\text{max}}\approx0$, we investigated how the trapping times, and thereby the effective quasi-stability of the ghost differs to that of the saddle.
For this, we divided each trajectory into N segments, explicitly integrated system Eqs.1 along each segment (between initial/final $(x_{\text{in,i}},y_{\text{in,i}})$/$(x_{\text{fin,i}},y_{\text{fin,i}})$ points; Supplementary section II), and determined analytically and numerically the local trapping times $\tau_i$ as a function of the local $\lambda_{\text{max}}^s$. The functional forms of the analytical expressions,
\[
  \tau_{\text{i,saddle}} =\frac{1}{\lambda_{\text{max}}^{sp}}\left[\ln \left|\frac{x-\lambda_{\text{max}}^{sp}/2}{x+\lambda_{\text{max}}^{sp}/2} \right|\right] \Biggr|_{x_{\text{in,i}}}^{x_{\text{fin,i}}}
\]
and,

\[
  \tau_{\text{i,ghost}} =\frac{2}{\lambda_{\text{max}}^{sp}}\left[\tan^{-1} \left(\frac{2x}{\lambda_{\text{max}}^{sp}}\right)\right] \Biggr|_{x_{\text{in,i}}}^{x_{\text{fin,i}}}
\]
and the corresponding numerical verification show that $\tau_i$ quickly decays along positive $\lambda_{\text{max}}^{sp}$ for the saddle, whereas for the ghost, a parabolic dependency on $\lambda_{\text{max}}^{sp}$ applies (Supplementary Fig.1(d)).
Contrary to the saddle, for which $\tau$ decreases monotonically with increasing $\sigma$, for the ghost, $\tau$ is robust with respect to noise\cite{Sardany_s_2020}, remaining constant over two orders of magnitude of $\sigma$ and decaying to half-maximum only at $\sigma \sim 10^{-1}$.
This characteristic of ghost sets is preserved for different $\alpha$ values (Supplementary Fig.1(e)). 

Following the features from this example, we now {\textit{define}} {\textit{ghost attracting sets}} formally. $\mathcal{A}$ is a ghost attracting set if {\textit {(a)}} it is a closed bounded set that does not contain any semi-trajectories in forward time; {\textit(b)} there is a closed set $\mathcal{B(A})$ with non-empty interior, the ghost basin of attraction, associated with $\mathcal{A}$, such that 
\begin{enumerate}
\item[(i)] for any   $x_0\in \mathcal{B}(\mathcal{A})$  there is a $t(x_0) \geq 0$ such that  $x(t;x_0)\in \mathcal{A}$ (attraction); 
\item[(ii)] for any   $x_0\in \mathcal{A}$  there is a $t(x_0)\leq 0$ such that  $x(t;x_0)\in \mathcal{B}(\mathcal{A})$ (minimallity); 
\item[(iii)] $\mathcal{A}$ is a proper subset of  $\mathcal{B}(\mathcal{A})$ (contraction).
\end{enumerate}
Moreover, $\mathcal{A}$ has entrance and exit boundaries with respect to the flow \cite{Tyukin_2024}.

\textit{Ghost channels and ghost cycles.}\textemdash 
We next extend the concept of ghost sets to complex ghost structures that we term \textit{ghost channels} and \textit{ghost cycles}, in analogy to heteroclinic structures constructed from saddles. Formally, we define a ghost channel as follows:

Let $\mathcal{A}_1$, $\dots$, $\mathcal{A}_N$ be N ghost attracting sets of the underlying system.  We say that the sets form a ghost channel if:
\begin{equation}
   \partial_{esc}{\mathcal{A}_i}\subset \mathcal{B}(\mathcal{A}_{i+1}), \ i=1,\dots,N-1, 
\end{equation}
where $\partial_{esc}{\mathcal{A}_i}$ is the escape boundary of $\mathcal{A}_i$ with respect to the flow. Thus, ghost channels (GChs) appear when multiple ghosts are aligned in a sequence such that the trajectory escaping a preceding ghost is directed by the flow through the next ghost in the sequence(Fig.~\ref{fig:fig1}(d),~\ref{fig:fig3}(b); Supplementary section III.B and Refs. [31-33]). Thus, GChs, in accordance to their definition, inherit attraction, whereas heteroclinic channels (HCh) may not. Comparing the reproducible guidance of trajectories through a GCh and a HCh (Fig.~\ref{fig:fig3}(a)), both constructed geometrically for simplicity, shows that only the GCh uniquely funnels the flow in phase-space, even for increased $\sigma$ (Fig.~\ref{fig:fig3}(b); Supplementary Fig.2(a-d)). In contrast, the trajectories stochastically exit along the HCh's unstable manifold. This shows that GChs, but not HChs, guarantee reproducible quasi-stable sequential switching dynamics, as also confirmed by the consistently lower Euclidean distance between the trajectories in the GCh (Fig.~\ref{fig:fig3}(c)). We find that GChs with equivalent properties are characteristic for a broad class of systems, including models of charge density waves, climate tipping cascades\cite{Dekker_2018}, and unidirectionally coupled cellular signaling model (Supplementary Fig.2(e), Section III.B). 

\begin{figure*}
\includegraphics[scale=1]{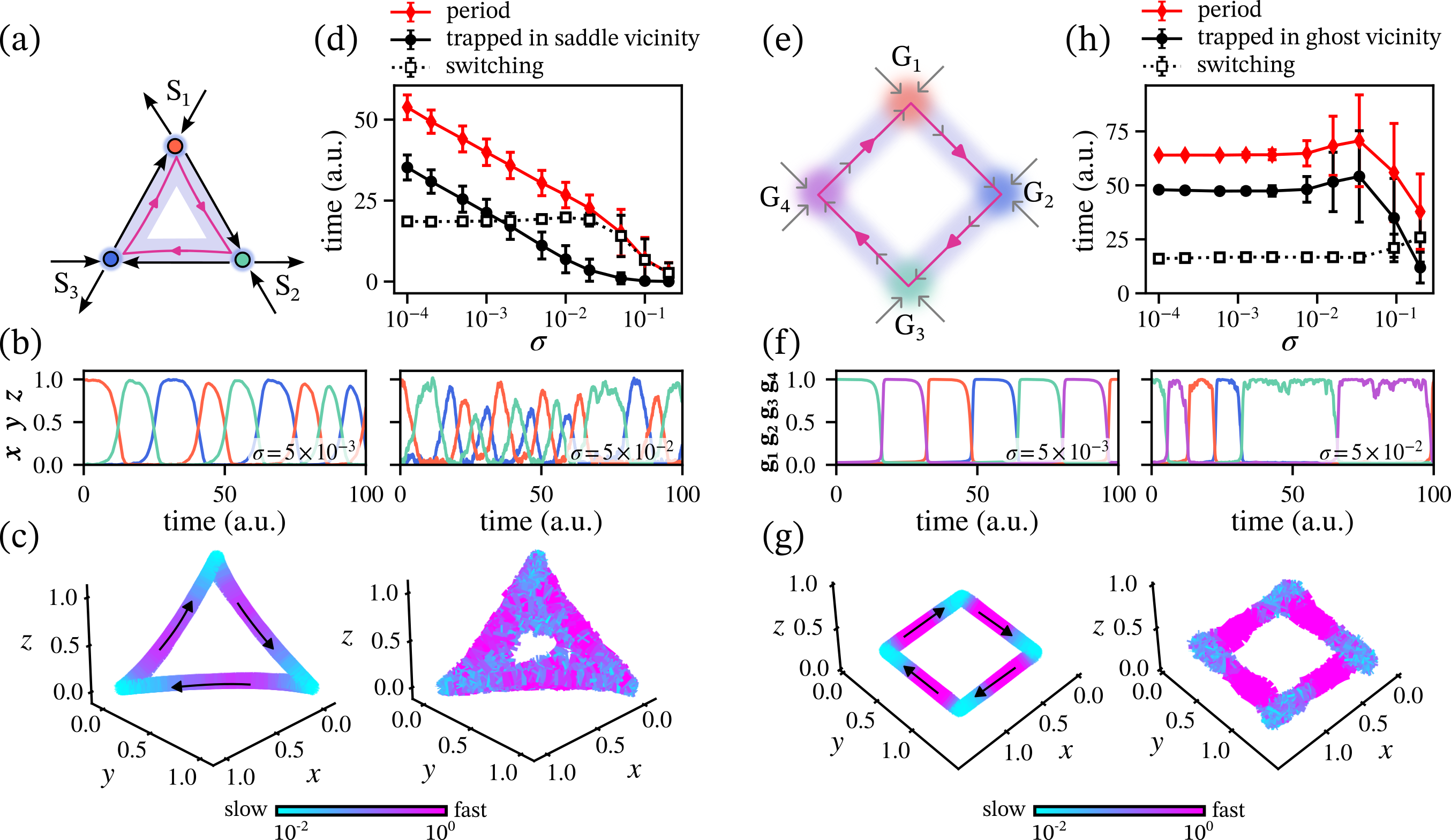}
\caption{\label{fig:fig4} Comparison of heteroclinic (HC) and ghost (GC) cycles. (a) Schematic of a HC of three saddles ($S_1$ to $S_3$) and (b) exemplary time-series for two noise intensities  $\sigma$. (c) Corresponding phase-space trajectories, color-coded by velocity. (d) Characteristic HC-times as a function of $\sigma$: HC period (red), total trapping time at the saddles (solid black) and  switching time (dashed black line). Vicinity was determined by three-dimensional spheres of radius $\epsilon = 0.1$ centered around the saddles. The mean $\pm$ root mean squared error of the s.d. over time is plotted from 30 trajectories. (e-g) Same as in (a-c), but for a GC (ghosts $G_1$ to $G_4$). (h) Characteristic GC-times as a function of $\sigma$. Labeling as in (d).}
\end{figure*}

To conceptualize the emergence of oscillatory quasi-stable sequential dynamics, we also constructed ghost cycles (GC; Fig.\ref{fig:fig4}(e)). Formally, if $\partial_{esc}{\mathcal{A}_N}\subset \mathcal{B}(\mathcal{A}_{1})$, we have a GC. To compare their dynamics to heteroclinic cycles (HC; Fig.\ref{fig:fig4}(a)), we matched the trapping times (in arbitrary units) along a single saddle and ghost to be similar at low $\sigma$ by adjusting the saddle-values of a generic noise-driven Lotka-Volterra HC model (Supplementary Fig.3). The period of the HC\cite{Rabinovich_2006}, $T\sim|ln \sigma| / \lambda_u$ decreases almost exponentially as $\sigma$ is increased, following the decrease of the total trapping time at the saddles (Fig. \ref{fig:fig4}(b),(d)). Moreover, the intervals in which the system's dynamics spends switching between the saddles within one HC period dominates already for intermediate noise $\sigma \leqslant 10^{-3}$. This is also reflected in the speed of the phase-space trajectories, which increasingly fill-in the phase-space regions distant to the heteroclinic backbone under increased noise (Fig.~\ref{fig:fig4}(c)). In contrast, increasing $\sigma$ does not affect the mean period of GCs over a large range of noise intensity, and the trajectories remain bounded along the cycle (Fig.~\ref{fig:fig4}(f-h)). The times spent on the ghosts remain $\sim$2-fold larger than the transition times between them, even for $\sigma$$>$$10^{-2}$. This is also reflected in the speed of the trajectory in phase-space, with a clear separation of the timescales (Fig.~\ref{fig:fig4}(g)). 
GCs thus uniquely provide a dynamical basis for emergence of robust and sustained quasi-stable oscillatory switching dynamics even for inherently noisy systems. In models, GCs can occur when a limit cycle terminates via a single or multiple saddle-node on invariant circle bifurcations (SNICs)\cite{Meeuse_2020,Farjami_2021,Sanchez_2022}. 
In the vicinity of the SNICs, the system's dynamics is governed by the SNs about to emerge: after being transiently trapped in the ghost set, the trajectory escapes and is trapped by the next one, thus continually switching between ghosts in the sequence (Supplementary Fig.5(a-c)). We demonstrate this also for a generic gene-regulatory network model proposed to underlie stem cell differentiation\cite{Farjami_2021} (Supplementary Fig.5(d-f)). 

\textit{Conclusions and outlook.}\textemdash We introduced ghost channels and ghost cycles as novel objects that guide flow in phase-space and give rise to reliable quasi-stable dynamics even in the presence of noise. This shows that quasi-stability can emerge in systems whose dynamics is neither organized by fixed points, nor dominated by limit cycle attractors (slow-fast systems \cite{Guckenheimer_1983,Fenichel1971PersistenceAS, Kuehn_2015}), the comparison to which we discussed elsewhere\cite{koch2024ghost}. We thus propose that ghost-based objects provide a possible mechanistic description for the emergence of ordered and reproducible transient behavior across living and man-made systems \cite{Mazor_2005, Kato_2015, Nichols_2017,Hastings_2018,Woo_2023, Colin_de_Verdi_re_2007,Benozzo_2021}. Using mainly geometric models to derive the basic definitions here, we identified the dynamical characteristics ($q\sim0$, eigenvalue gradient and escape paths), which could be used in an algorithmic fashion to identify ghost sets in any arbitrary-dimensional complex system.

The proposed concepts could potentially provide a way forward in identifying mechanisms e.g. of emergence of  complex low-dimensional manifolds\cite{Langdon_2023} typical for neuronal activity data during cognitive or behavioral tasks\cite{Kato_2015,Nichols_2017,Morrison_2022}, as it is easy to imagine hybrid structures of ghosts and saddles giving rise to new phase-space objects benefiting from different properties (Supplementary Fig.6). Moreover, the presence of distinct time-scales emerging from the ghost scaffolds could aid development of time-series analysis methods for detecting quasi-stable patterns and corresponding transitions, e.g. via phase-space-based metrics\cite{beimGraben2013}. Our conceptual framework therefore provides new perspectives on natural systems where long transients are common.

We thank A. Aulehla, K. Lehnertz and P. Francois for valuable feedback on the manuscript, and J. Gunawardena and J. Garcia-Ojalvo for insightful discussions. A.K. acknowledges funding by the Lise Meitner Excellence Programme of the Max Planck Society, D.K. - EMBO Fellowship (Grant nr. ALTF 310-2021), I.T. -  UKRI Turing AI Fellowship EP/V025295/2.


\providecommand{\noopsort}[1]{}\providecommand{\singleletter}[1]{#1}%

\renewcommand{\figurename}{Supplementary Fig.}
\counterwithin{figure}{section}
\renewcommand\thefigure{\arabic{figure}}

\newpage
\section*{Supplementary Materials}


\section{Numerical estimation of eigenvalue spectrum at phase space region of slow dynamics}

Generally, in the regions where the norm of the dynamics is close to zero ($q({\bf x})<q_{\text{thresh}}$), local linear expansion of the dynamics is still valid.
When linearizing around a slow point, ${\bf x^{sp}}$, the local linear system
takes the form,
\setcounter{equation}{1}
\begin{eqnarray}
\begin{aligned}
\dfrac{{\bf d\delta x}}{dt}={\bf F(x^{sp})}+ {\bf F'(x^{sp}) \delta x}
\label{eq:six}
\end{aligned}
\end{eqnarray}

where ${\bf F'(x^{sp})}$ is the Jacobian of the system evaluated at the slow point, and ${\bf \delta x}$ is the perturbation added to ${\bf x^{sp}}$. In practice, it can be considered that the constant term ${\bf F(x^{sp})}$ is negligible, therefore it can be set to zero \cite{Sussillo2013}. This enables to estimate the eigenvalues of ${\bf F'(x^{sp})}$, $\lambda_{\text{min}}^{sp}$ and $\lambda_{\text{max}}^{sp}$ as in classical linear stability analysis \cite{strogatz:1994}, and approximate the local dynamics of the slow points using the corresponding eigenvalues. 

\section{Analytical description of trapping time at ghost/saddle states}

In order to find the dependence between the trapping time at the ghost state and the local eigenvalues, let us consider the normal form of the saddle-node bifurcation given in system Eqs. 1 (Main text). The system has steady states for $x^{*}=\pm\sqrt{-\alpha}$ (a stable fixed point and a saddle), and a saddle-node bifurcation at $\alpha=0$. The saddle value in Fig.2 (main text): $\nu=\frac{Re \lambda_s}{\lambda_u}\approx 1.11$,
$\lambda_s$ and $\lambda_u$ are the stable and unstable eigenvalues respectively, $Re$ stands for the real part. The Jacobian of the system is given by,

\begin{eqnarray}
J=\begin{bmatrix}
2x^{*} & 0 \\
0 & -1
\end{bmatrix}
\label{eq:two}
\end{eqnarray}
 with eigenvalues $\lambda_{\text{min}}=-1$ and $\lambda_{\text{max}}=2x^*$. 

Since system Eqs.1 are explicitly integrable, the time of flight of the trajectory $\tau(\alpha)$ from an initial $(x_{\text{in}},y_{\text{in}})$ to a final point $(x_{\text{fin}}, y_{\text{fin}})$ is given by,

\begin{eqnarray}
\begin{aligned}
\tau(\alpha)&=\int_{x_{\text{in}}}^{x_{\text{fin}}} \frac{1}{dx/dt} \,dx\\
&= \int_{x_{\text{in}}}^{x_{\text{fin}}} \frac{1}{\alpha+x^2} \,dx\\
\label{eq:three}
\end{aligned}
\end{eqnarray}

For $\alpha<0$ (where the saddle and the stable fixed point co-exist) this integral is given by,

\begin{eqnarray}
\begin{aligned}
\tau(\alpha)&= \int_{x_{\text{in}}}^{x_{\text{fin}}} \frac{1}{-|\alpha|+x^2} \,dx\\
&= \frac{1}{2\sqrt{|\alpha|}}(\text{ln}|\frac{x_{\text{fin}}-\sqrt{|\alpha|}}{x_{\text{fin}}+\sqrt{|\alpha|}}|-\text{ln}|\frac{x_{\text{in}}-\sqrt{|\alpha|}}{x_{\text{in}}+\sqrt{|\alpha|}}|)
\label{eq:four}
\end{aligned}
\end{eqnarray}

For $\alpha>0$ the integral is given by:

\begin{eqnarray}
\begin{aligned}
\tau(\alpha)&= \int_{x_{\text{in}}}^{x_{\text{fin}}} \frac{1}{\alpha+x^2} \,dx\\
&= \frac{1}{\sqrt{\alpha}}(\text{tan}^{-1}(\frac{x_{\text{fin}}}{\sqrt{\alpha}})-\text{tan}^{-1}(\frac{x_{\text{in}}}{\sqrt{\alpha}}))
\label{eq:four_3}
\end{aligned}
\end{eqnarray}

Since a ghost state exists for $\alpha \rightarrow 0^+$, the total trapping time along the trajectory in the region of slow points can be analytically estimated as a sum of local trapping times calculated by integrating along each of the trajectory's segments (the trajectory is divided into $N$ segments of length $2w = 0.01$, Supplementary Fig. 1(c)):

\begin{eqnarray}
\begin{aligned}
\tau_i(\alpha)&= \frac{1}{\sqrt{\alpha}}(\text{tan}^{-1}(\frac{x_{\text{fin,i}}}{\sqrt{\alpha}})-\text{tan}^{-1}(\frac{x_{\text{in,i}}}{\sqrt{\alpha}}))
\label{eq:four_1}
\end{aligned}
\end{eqnarray}
where $i \in [0,N]$ denotes the $i^{th}$ piece of trajectory between $x_{\text{in,i}},x_{\text{fin,i}}$ with $x_{\text{in,0}}=x_{\text{in}}$, $x_{\text{fin,N}}=x_{\text{fin}}$ for the boundary segments and $x_{\text{in,i}}=x_{\text{fin,(i-1)}}$ for the rest. The total trapping time is then given by $\sum_{i=1}^{N} \tau_i(\alpha) = \tau(\alpha)$.

Let us consider a fictitious fixed point $(x^f,y^f)=(\sqrt{\alpha},0)$ which is a reflection of the saddle fixed point that disappeared at $\alpha=0$. In the vicinity of the bifurcation this fictitious fixed point satisfies the criteria of a slow point ($q(x^f,y^f)<q_{\text{thresh}}$), hence the local eigenvalue along the ghost can be approximated to be $\lambda_{\text{max}}^{sp}=\lambda_{\text{max}}^i=2\sqrt{\alpha}$. Using this relation, an expression of $\tau_i$ as a function of the eigenvalue $\lambda_{\text{max}}^i$ is given by:

\begin{eqnarray}
\begin{aligned}
\tau_i(\lambda_{\text{max}}^i)&= \frac{2}{\lambda_{\text{max}}^i}(\text{tan}^{-1}(\frac{2x_{\text{fin,i}}}{\lambda_{\text{max}}^i})-\text{tan}^{-1}(\frac{2x_{\text{in,i}}}{\lambda_{\text{max}}^i}))
\label{eq:five}
\end{aligned}
\end{eqnarray}

Numerical and analytical dependence of the local trapping time on maximal eigenvalue is shown in Supplementary Fig. 2(d). Given that the trapping time is a local quantity, the approximation is valid only for small $w$ ($x_{\text{fin,i}} \approx x_{\text{in,i}}$). 

Similarly, from Eq.~(\ref{eq:four}) the local trapping time around the saddle fixed point is given by:

\begin{eqnarray}
\begin{aligned}
\tau_i(\lambda_{\text{max}}^i)&=\\ &\frac{1}{\lambda_{\text{max}}^i}(\text{ln}|\frac{x_{\text{fin,i}}-\lambda_{\text{max}}^i/2}{x_{\text{fin,i}}+\lambda_{\text{max}}^i/2}|-\text{ln}|\frac{x_{\text{in,i}}-\lambda_{\text{max}}^i/2}{x_{\text{in,i}}+\lambda_{\text{max}}^i/2}|)
\label{eq:four_2}
\end{aligned}
\end{eqnarray}

For the estimation of total trapping time in Fig. 2(c) and Supplementary Fig. 1(e), initial conditions are fixed (separately for saddle fixed point and ghost state) for the 30 repetitions.

\section{Model equations and parameters}

\subsection{Construction of generic heteroclinic/ghost channels (HCh/GCh) and ghost cycles (GC)}\label{sec:genModelling}

To define ghost channels/cycles as phase space objects, we used a generic, geometric modeling approach to define the flow in an abstract 2D/3D phase space, following the method described by Morrison \& Young\cite{Morrison_2022} (a similar approach has been introduced by Corson and Siggia\cite{Corson_2012}). For this, the phase space is partitioned into $1\leq i \leq k$ distinct areas or volumes, and for each area or volume, a function $f_i$ is chosen that defines specific dynamics in each partition (e.g. ghost, saddle, uniform flow etc.). The full system's dynamics is thus given by $\frac{d}{dt}\bf{x} = \sum^n_{i=1} w_i(\bf{x})f_i(\bf{x})$, where $w_i(\bf{x})$ assigns a high weight to $f_i$ if and only if $\bf{x}$ is within the $i^{th}$ partition of the phase space, ensuring a unique dynamics. Specifically, partitions of the phase space are given by cubes defined by coordinates $x_i^{\text{min}}$,$x_i^{\text{max}}$,$y_i^{\text{min}}$, $y_i^{\text{max}}$, $z_i^{\text{min}}$ and $z_i^{\text{max}}$ and weighting functions by

\begin{eqnarray*}
\begin{aligned}
w_i(\bf{x}) = \frac{1}{4}&(\text{tanh}(\gamma(x-x_i^{\text{min}})) - \text{tanh}(\gamma(x-x_i^{\text{max}})))
\\\times&(\text{tanh}(\gamma(y-y_i^{\text{min}})) - \text{tanh}(\gamma(y-y_i^{\text{max}})))
\\\times&(\text{tanh}(\gamma(z-z_i^{\text{min}})) - \text{tanh}(\gamma(z-z_i^{\text{max}}))),
\end{aligned}
\end{eqnarray*}

\noindent where $\bf{x} = (x,y,z)^T$ and $\gamma$ defines how steep the transitions in the weighting functions between partitions are. Generally, the dynamics within a partition is defined by $f(\bf{x}) = (\hat{f}(x),\hat{g}(y), \hat{h}(z))$, where $\hat{h}(z) = -(z-0.5)$. For ghosts dynamics, $\hat{f}$ is given by $x \mapsto c_i(\alpha + (x-x_{\text{offset}})^2)$, $c_i \in (-1,1)$, and $\hat{g}(y)$ by $y \mapsto \tilde{c_i}(y-y_{\text{offset}})$, $\tilde{c_i} \in (-1,1)$, or vice versa. The offset defines the center of the partition. In both functions, $c$ and $\tilde{c}$ were chosen to funnel the flow from three directions on the x/y plane. Saddle dynamics is defined by $\hat{f}(x)$ given by $x\mapsto -\nu(x-x_{\text{offset}})$ defining the stable manifold, and $\hat{g}(y)$ by $y \mapsto (y-y_{\text{offset}})$ defining the unstable manifold, or vice versa. $\nu$ is the saddle value. The uniform flow is defined by $\hat{f}(x) = m_1$, $\hat{g}(y) = m_2$, $m_1,m_2 \in \mathbb{R}$. For 2D systems, the $z$ dimension (and thus $\hat{h}$) was omitted.

To construct a heteroclinic channel in 2D, the functions $\mathbf{f_i}$ and $\mathbf{w_i}$ were chosen such that four saddles are positioned within partitions on a diagonal and the flow on the neighbouring partitions was defined to be uniform (Fig. 3(a) and Supplementary Fig. 2(a)). To construct a ghost channel in 2D, the functions $\mathbf{f_i}$ and $\mathbf{w_i}$ were chosen such that the flow from the preceding ghost is directed towards the following ghost in the sequence (Fig. 1(d), and 3(b) and Supplementary Fig. 2(c)). For a ghost cycle, the flow of the last ghost in the sequence is directed towards the first ghost (Fig. 4(e)). The full set of equations and parameters for these systems can be found in the {\tt Python} and ${\tt XPPAUT}$\cite{Ermentrout_2002} code accompanying this article.

To obtain similar shape of the ghost cycle oscillations (Supplementary Fig. 3(a)) to that of the heteroclinic cycle oscillations (cf. Eqs. \ref{eq:horchler} and Fig. 4(a,b)), a Euclidean distance of each point in the trajectory to each of the ghost position was obtained ($d_j = || \overline{x} - G_j ||, 1\leq j \leq 4$, Supplementary Fig. 3(b) left), and the corresponding distance time series (Supplementary Fig. 3(b), right) were mapped using a Hill-type function $\Theta(d) = d^{-3}/(d^{-3} + 0.3^{-3})$ (Supplementary Fig. 3(c)). This ensures that the shape of the oscillations in the ghost cycle are comparable to those in the heteroclinic cycle.  To construct HC and GC that have comparable trapping time on the saddles/ghosts, the saddle value was set to $\nu=4$ to adjust the differences of the average time spent in saddle/ghost vicinity (colored triangles) to $< 1\%$ for the lowest noise level used in this work ($\sigma = 10^{-4}$, Supplementary Fig. 3(d)).

To quantify the robustness of the trajectories within the heteroclinic and ghost channels, we estimated the Euclidean distance (ED) between pairs of trajectories from different realizations of stochastic simulations: $ED=\sqrt{(x_2(t)-x_1(t))^2+(y_2(t)-y_1(t))^2}$, where $1,2$ represent trajectory pairs. To avoid effects from different trapping times on the ghosts/saddles, the trajectories were initially time warped using the \href{https://pypi.org/project/dtaidistance/}{{\tt dtaidistance}} package (v 2.3.9) in {\tt Python}. The average ED over time ($t_{\text{end}} = 500$ a.u.) of trajectories starting from 6 different initial conditions with 30 replicates each (giving a total of 180 trajectories) (mean $\pm$ root mean squared error of the standard deviation over time) is shown in Fig. 3(c).

\subsection{Supplementary examples of ghost channels}
An infinite number of saddle-node bifurcations are characteristic for a model for charge-density waves\cite{Strogatz_1989}. Parameterizing the system in the vicinity of the SN bifurcations results in the formation of a ghost channel:

\begin{equation}
\begin{split}
\dot{x}&= E + 0.5\left(\text{sin}(x+y) + \text{sin}(x-y)\right)\\
\dot{y}&= \frac{1}{1+ 2\gamma}\left(-2Kxy +0.5(\text{sin}(x+y) + (x-y))\right)\\
\end{split}
\label{eq:nghost}
\end{equation}

where E is proportional to the applied electric field, $\gamma$ is the viscous coupling strength and K - the elastic coupling strength. For E=0.9, $\gamma=0.5$, K=0.9, a ghost channel is observed. Phase-space with numerical $\lambda_{\text{max}}^{sp}$ estimation at the regions of slow dynamics and corresponding time series are shown in Supplementary Fig. 2(e), left. 

We next considered a climate model of cascading tipping, i.e. a sequence of abrupt transitions \cite{Dekker_2018}:

\begin{equation}
\begin{split}
\dot{x}&= a_{1}x^3 + a_{2}x + \phi\\
\dot{y}&= b_{1}x^3 + b_{2}x + 0.48x\\
\end{split}
\label{eq:climate}
\end{equation}

For $a_1=-0.5, a_2=0.5, b_1=-0.5, b_2=1, \phi=0.2$, the system exhibits a ghost channel. Corresponding phase-space with numerical $\lambda_{\text{max}}^{sp}$ estimation at the regions of slow dynamics and respective time-series are shown in Supplementary Fig. 2(e), middle.

As a third example, we adapted a bistable biochemical model from Ferrel \textit{et al}. \cite{ferrell2014ultrasensitivity} to construct a GCh by unidirectionally coupling two biochemical modules:

\begin{eqnarray}
\begin{aligned}
\dot{x} &= \left( k_{1}S + \frac{k_{2}x^n}{(K^n+x^n))}\right)(x_{\text{tot}}-x)-k_{3}x\\
\dot{y} &= \left((k_{1}x + \frac{k_{2}y^n}{(K^n+y^n))}\right)(x_{\text{tot}}-y)-k_{3}y\\
\end{aligned}
\label{eq:eight}
\end{eqnarray}
For $S = 0, x_{\text{tot}} = 1, k_{1} = 1, k_{2} = 2, k_{3}= 1, K = 0.505, n = 4$, the model has two ghost states in a sequence (ghost channel) followed by a stable fixed point. The corresponding phase space with the numerical $\lambda_{\text{max}}^{sp}$ estimation at the regions of slow dynamics, and the time series are shown in Supplementary Fig. 2(e) right.

\subsection{Model of a heteroclinic cycle (HC)}

To model heteroclinic cycle dynamics, we used the 3D HC described by Horchler \textit{et al}. \cite{Horchler_2015}:

\begin{equation}
\frac{d}{dt} \textbf{x} = \textbf{x}\cdot(\alpha - \rho\textbf{x}),
\label{eq:horchler}
\end{equation}

\noindent where $\textbf{x}=(x,y,z)^T$, $\alpha = (\alpha_1, \alpha_2, \alpha_3)$, \\

\noindent $\rho =
\begin{bmatrix}
\alpha_1/\beta_1 & (\alpha_1+\alpha_2)/\beta_2 & (\alpha_1 - \frac{\alpha_3}{\nu_3})/\beta_3\\
(\alpha_2 - \frac{\alpha_1}{\nu_1})/\beta_1 & \alpha_2/\beta_2  & (\alpha_2+\alpha_3)/\beta_3 \\
(\alpha_3+\alpha_1)/\beta_1 & (\alpha_3 - \frac{\alpha_2}{\nu_2})/\beta_2 & \alpha_3/\beta_3 
\end{bmatrix}$,\\

\noindent $\alpha_i = 2, \beta_i = 1$ and $\nu_i = 4$, for $i=1,..3$.

\subsection{Supplementary example of a ghost cycle}

We analyzed additionally a genetic network model described in\cite{Farjami_2021}:
\begin{eqnarray}
\begin{aligned}
\dot{x} &=  b + \frac{g}{(1+\alpha(y^h))(1+\beta(z^h))} - d x\\
\dot{y} &=  b + \frac{g}{(1+\alpha(z^h))(1+\beta(x^h))} - d y\\
\dot{z} &=  b + \frac{g}{(1+\alpha(x^h))(1+\beta(y^h))} - d z\\
\end{aligned}
\label{eq:gcycle}
\end{eqnarray}

For $\alpha=9$, $\beta=0.1$, $h=3$, $d=0.2$, $b = 10^{-5}$ and $g=1.51$, the limit cycle oscillations terminate at three SNIC bifurcations (Supplementary Fig. 5(d)). For parameter organization before the SNIC, the behavior of the system is guided by the three SN that are about to appear, effectively displaying ghost cycle oscillations. The bifurcation diagram, the phase trajectories (colour-coded by the speed) for different parameter values, as well as the emergence of a stable fixed point and a saddle right after the SNIC bifurcation are shown in Supplementary Fig. 5(d)-(f).

\section{Model of noise-induced attractor switching}
To model classical 'metastability' arising from noise-induced switching between fixed points with shallow basins of attraction\cite{Bovier_2015, Rossi2023}, we used the model of a ghost cycle described in Supplementary Section III.A, parametrized after the SNICs ($\alpha=-0.002$). Thus, the results shown in Supplementary Fig.4 correspond to four metastable attractors in the above sense. This model also fails to ensure reliable quasi-stability. In this case, high noise intensity is required to induce the switching, whereas the period displays high sensitivity to noise increments. 

\section{Model of a hybrid ghost-saddle phase-space object}

The hybrid ghost-saddle structure shown in Supplementary Fig. \ref{fig:SFig_hybrid} was generated using the geometric approach described in Supplementary Section \ref{sec:genModelling}. Briefly, the phase space was designed to feature two adjacent ghost cycles with counterclockwise rotations which meet and intersect at a saddle point were the presence of noise allows for switching between the orbits.  For ghost elements, we used $\alpha=0.002$, whereas the saddle was given a saddle value $\nu = 1.4$. Bifurcation analysis was performed to study the dynamic mechanism by which the object can emerge.

\section{Model implementation, numerical integration and characterization of the dynamics}

The ODE models were integrated using the 4th order Runge-Kutta scheme implemented using custom-made {\textit{Python}} code. The contribution of the noise is modeled as a Wiener process where Gaussian white noise is introduced as an additive term at each time step. This results in a stochastic differential equation (SDE) in $It\hat{o}$ form, $\dot{X(t)} = f(X(t), t)dt + \sigma dW(t)$, where $dt$ denotes the step size (0.05 or 0.01 were used throughout this study), where $\sigma(t)$ describes the additive noise intensity and $W(t)$ denotes a Wiener process whose independent increments follow a normal distribution with $(\mu=0,var=\sqrt{dt})$.  
For simulation of stable heteroclinic cycles, integration of the system was performed by restricting $\textbf{x}$ to $\mathbb{R}^{+}$ as described in \cite{Horchler_2015}, to avoid escaping from the heteroclinic cycle via the unstable directions.
The velocity of the trajectories in phase space was normalized between the 5-95 percentile of the velocity values along the trajectories for the different noise intensities form the SC and the GC independently.

All codes for reproducing the results and figures from this study are available on GitHub at \url{https://github.com/KochLabCode/BeyondFixedPoints}.

\section{Supplementary References}
\bibliography{aapmsamp}

\newpage
\onecolumngrid
\section{Supplementary Figures}
\begin{figure*}[!h]
\includegraphics[scale=1]{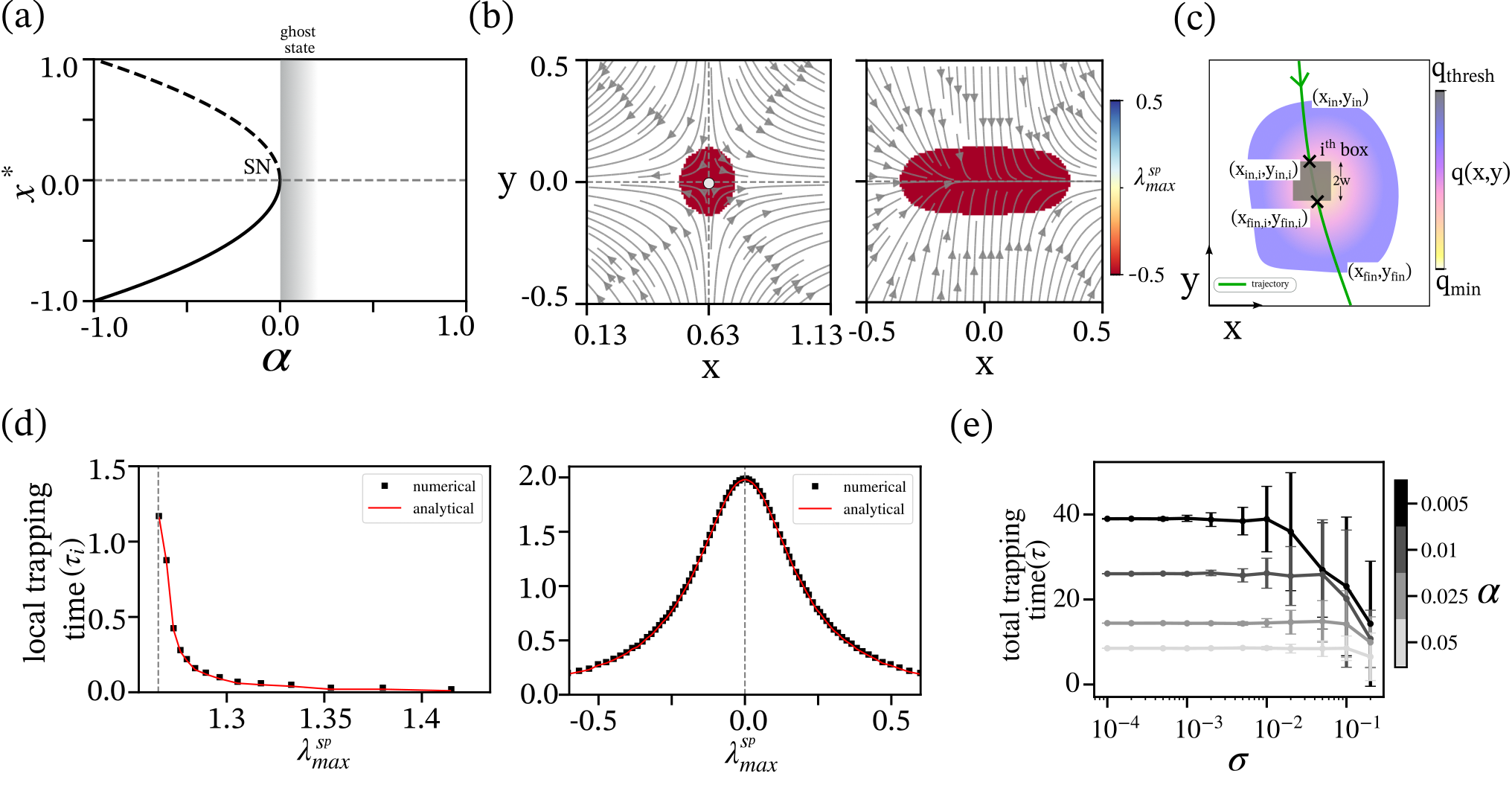}
\caption{Characterization of the dynamics of saddle fixed points and ghost states. (a) Bifurcation diagram corresponding to Eq. 1 (Main text). Solid/dashed lines: stable/unstable fixed points. Gray shaded area: region of ghost state. SN: saddle-node bifurcation. (b) $\lambda_{min}^{sp}$ at the region of slow dynamics for the saddle fixed point (left) and the ghost state (right). Corresponding to Fig. 2(a),(b). (c) Schematic of the total/local trapping time estimation in the region of slow dynamics. Black crosses: entry and exit points of trajectory into the $i^{th}$ box, ($x_{in},y_{in}$) and ($x_{fin},y_{fin}$) are the entry and exit points of trajectory in the region. (d) Analytical and numerical piece-wise trapping time (cf. Supplementary section II) as a function of $\lambda_{\text{max}}^{sp}$ for a trajectory transversing across the $q(x,y)<q_{thresh}$ region of the saddle (left) and the ghost state (right). Dashed line: $\lambda_{max}^{sp}$ where $q(x,y)$ is minimum. (e) Total trapping time as a function of noise intensity $\sigma$ for the model (Eq. 1, Main text) for different $\alpha$. Mean $\pm$ s.d. from 30 realizations starting from different initial conditions are shown.}
\label{fig:fig2_supp} 
\end{figure*}

\begin{figure*}
\includegraphics[scale=0.95]{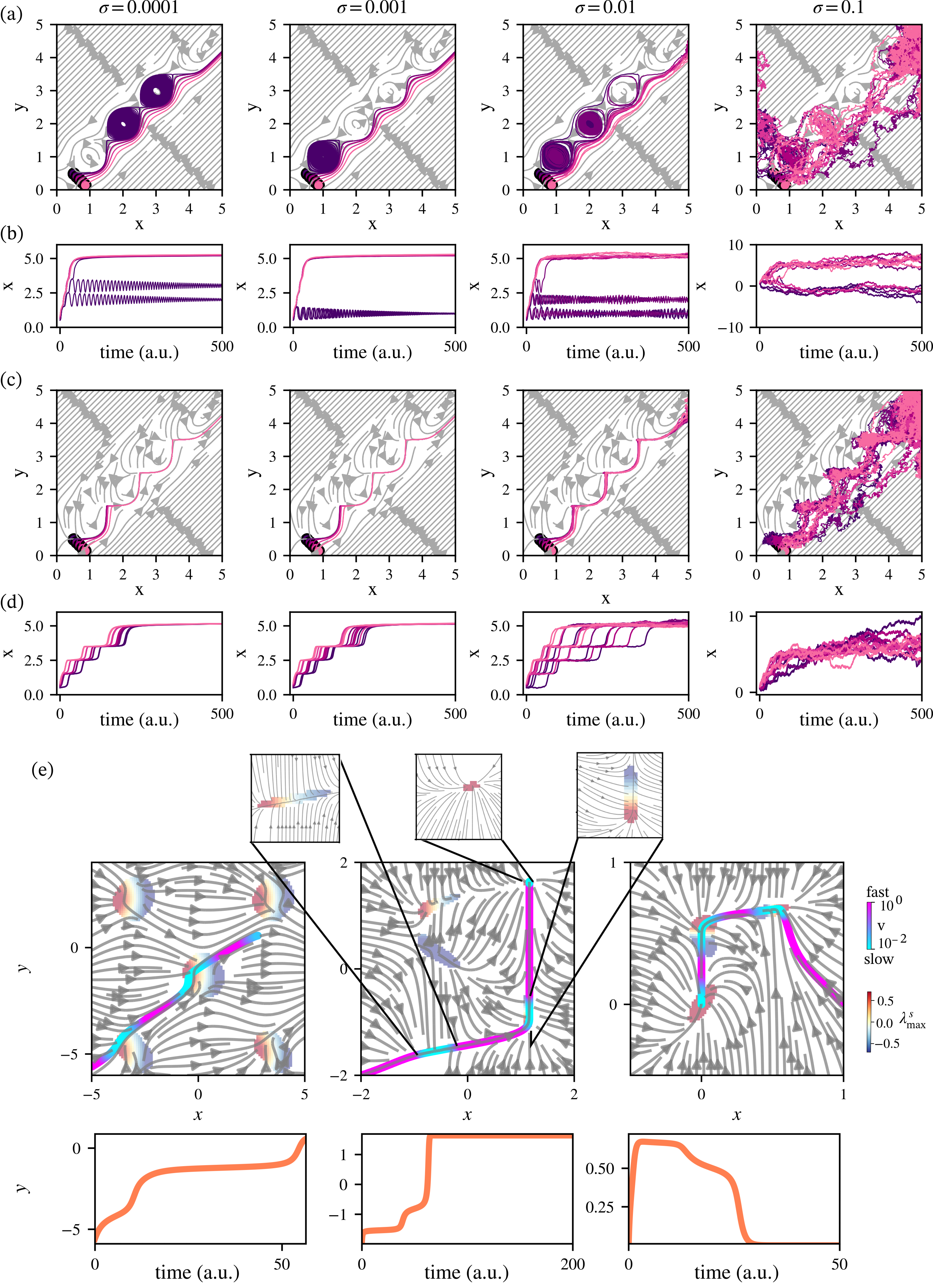}
\newpage
\caption{\label{fig:SFig_GCraw} Dynamics of heteroclinic (HCh) and ghost (GCh) channels. (a) Exemplary phase space trajectories from different initial conditions (3 repeats per initial condition) and for different $\sigma$ for a heteroclinic channel. Corresponding to Fig. 3(a). (b) Corresponding time series of the sample trajectories. (c), (d) Same as in (a), (b) but for a ghost channel. Corresponding to Fig. 3(b). See Supplementary materials section III.A for more details. (e) Top: Phase space trajectories (color-coded by the velocity) and the numerical $\lambda_{\text{max}}^{sp}$ estimation at the regions of slow dynamics for models Eqs. \ref{eq:nghost}, \ref{eq:climate} and \ref{eq:eight} ($q_{thresh}=0.1$, $0.02$, $0.01$) from left to right. Bottom: corresponding time series.}
\end{figure*}

\begin{figure*}
\includegraphics[scale=0.8]{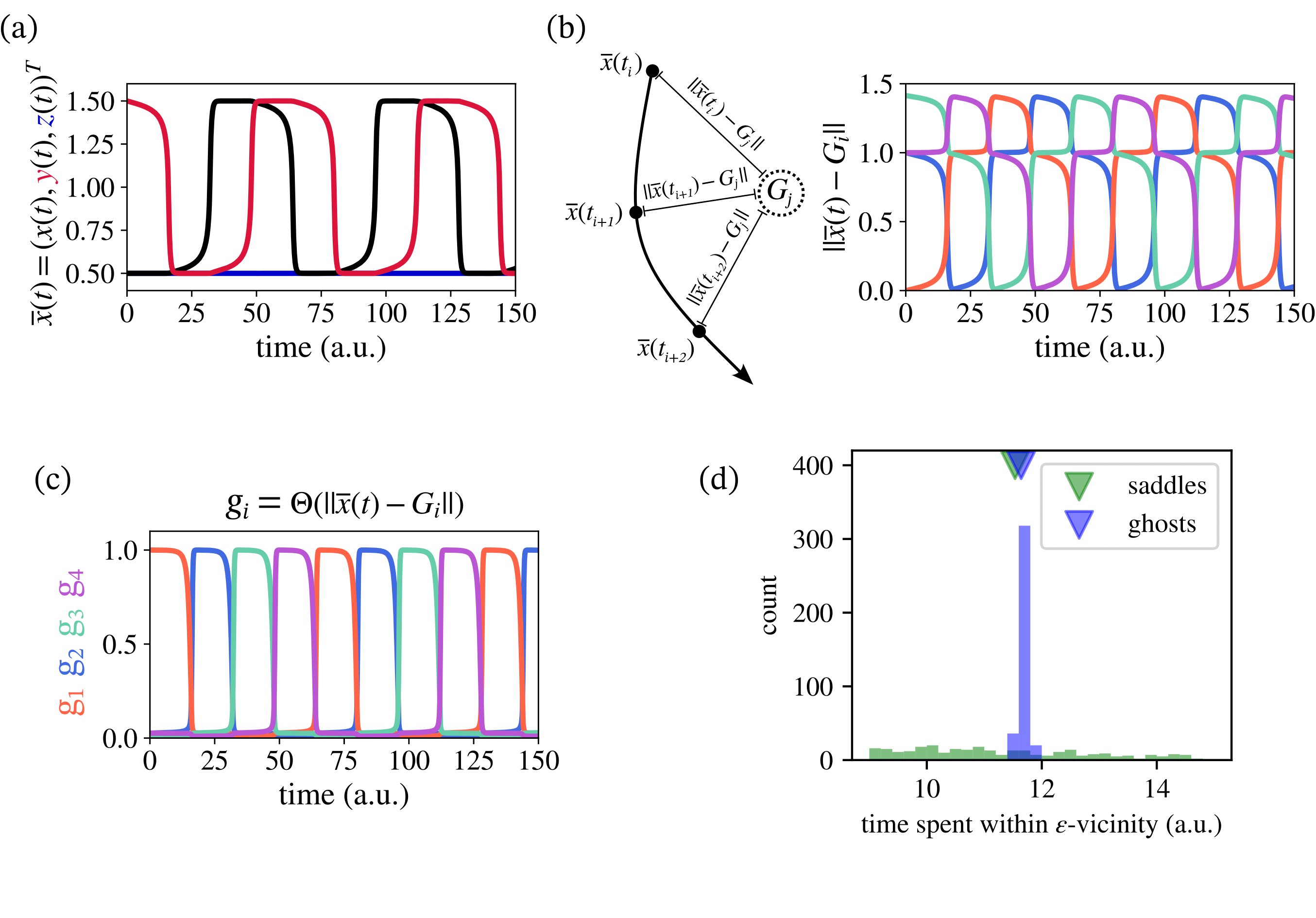}
\caption{\label{fig:SFig_hill} Characteristics of the shape of GC/HC oscillations and trapping times. (a) Time courses of the system variables for the ghost cycle corresponding to Fig. 4(e). (b) Schematic and exemplary time series of Euclidean distances between each point in the trajectory to each of the ghost position. (c) Mapped distance time series using a Hill-type function $\Theta$. (d) To conduct the comparative analyses between heteroclinic and ghost cycles starting from a similar baseline, the times spent within the vicinity of the ghost/saddle points were calculated and the saddle value was adjusted to minimize the differences of the average time spent in saddle/ghost vicinity (colored triangles) for the lowest noise level used in this work ($\sigma = 10^{-4}$). At $\nu=4$, the difference between average trapping in saddle/ghost states reached $< 1\%$, which was thus used for all simulations of the HC corresponding to Fig. 4(a). See also Supplementary materials, section III.A}
\end{figure*}

\begin{figure*}[!h]
\includegraphics[scale=1]{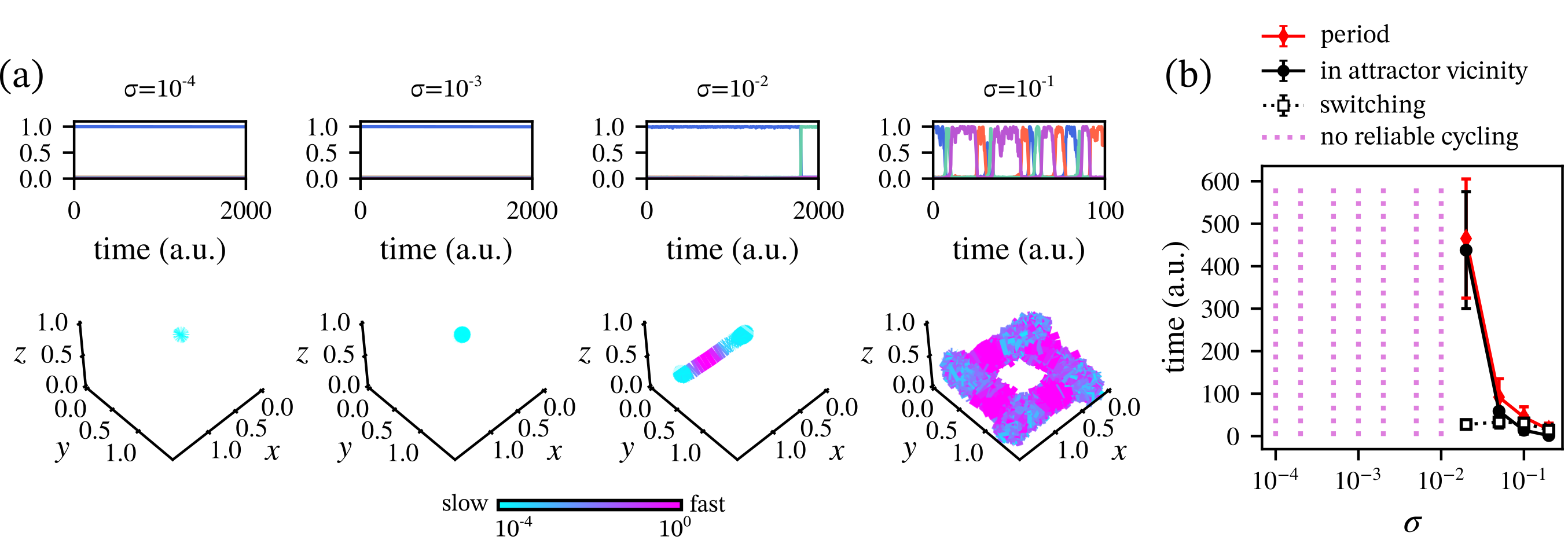}
\caption{\label{fig:SFig_hybrid} 
Characterizing metastability from noise-induced swithcing between stable fixed points. (a) Time-series and corresponding phase-space trajectories color coded by the speed of the trajectory for different noise intensities for the model of the ghost cycle described in section III.A in Supplementary materials, parameterized after the SNIC bifurcation (Section IV), $\alpha=-0.002$. (b) Characteristic times as a function of $\sigma$:  period (red), total trapping time in the stable fixed points (solid black) and switching time (dashed black line). A period is defined by completing at least one full cycle within $t = 2000$ arbitrary time units in all of the n = 30 replicates (the mean $\pm$ root mean squared error of the s.d. over time is plotted from 30 trajectories). Vicinity was determined by three-dimensional spheres of radius $\epsilon = 0.1$ centered around the middle between saddle and fixed points. See description in Supplementary Section IV.}
\end{figure*}

\begin{figure*}
\includegraphics[scale=1]{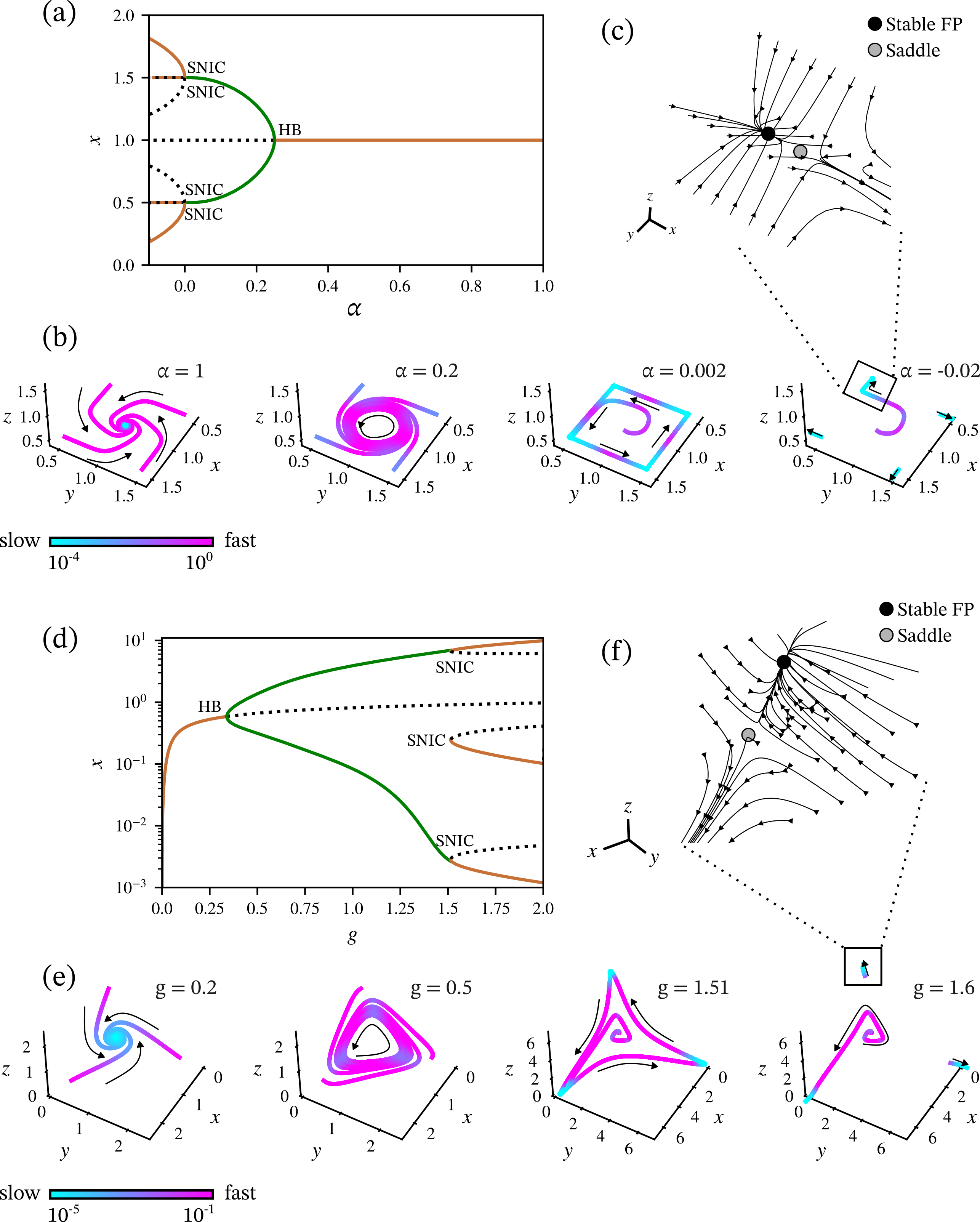}
\caption{\label{fig:SFig_SNIC} Mechanism of emergence of a ghost cycle. (a) Bifurcation diagram of ghost cycle described in section III.A in Supplementary materials. Brown/green lines: stable fixed point/limit cycle, dashed line: saddle fixed point. HB: Andronov-Hopf bifurcation, SNIC: saddle-node on invariant cycle bifurcation. (b) Phase space trajectories for different parameter values depicting the different dynamical regimes in (a), colour-coded by the speed in phase space. (c) Phase-space cut-out depicting the emergence of a stable fixed point and a saddle after the SNIC bifurcation. (d) Equivalent as in (a), only for the supplementary ghost cycle model, Eq. \ref{eq:gcycle}. Line description and notation as in (a). (e) Corresponding phase-space trajectories for different parameter values depicting the different dynamical regimes in (d). (f) Same as in (c), only for the model in (d).}
\end{figure*}

\newpage

\begin{figure*}
\includegraphics[scale=1]{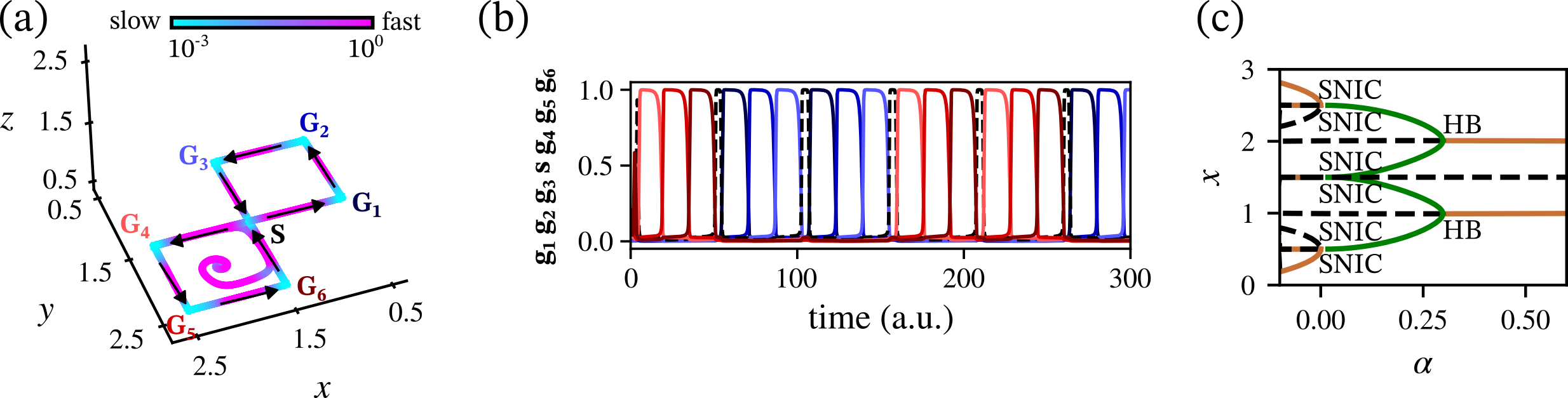}
\caption{\label{fig:SFig_hybrid} Hybrid phase-space object. (a) Two GCs joined by a saddle fixed point ($S$). (b) Corresponding time-courses for $\sigma = 10^{-4}$. (c) Respective bifurcation diagram. The dynamics in (b) emerges for $\alpha \approx 0^{+}$ (vicinity of the SNICs). Brown/green lines: stable spiral-fixed-points/limit cycles; dashed: unstable saddles. HB: Hopf bifurcation. See Supplementary Section V.}
\end{figure*}

\end{document}